\documentclass[prb,pre,aps,twocolumn,amsmath,amssymb,floatfix,superscriptaddress]{revtex4-2}
\usepackage{physics,graphicx,hyperref}
\hypersetup{colorlinks=true,citecolor=blue,linkcolor=blue,urlcolor=blue}
\usepackage[utf8]{inputenc}
\usepackage{mathtools}

\begin{document}

\title{Topological flat bands and higher-order topology in square-octagon lattice}

\author{Amrita Mukherjee}
\affiliation{Department of Condensed Matter Physics and Materials Science, Tata Institute of Fundamental Research, Mumbai 400005, India}

\author{Bahadur Singh}
\email{bahadur.singh@tifr.res.in}
\affiliation{Department of Condensed Matter Physics and Materials Science, Tata Institute of Fundamental Research, Mumbai 400005, India}

\begin{abstract}

Extensive recent research on Lieb and kagome lattices highlights their unique physics characterized by the coexistence of Dirac points, van Hove singularities, and flat bands. In these models, flat bands are typically pinned at the center of the Lieb spectrum or the extrema of kagome bands, offering limited tunability. In this work, we investigate the square-octagon lattice and demonstrate that flat bands generated through next-nearest-neighbor (NNN) hoppings can be tuned with intercell hoppings or staggered magnetic fluxes. Importantly, the introduction of staggered magnetic fluxes leads to the emergence of a Chern insulator phase and a higher-order topological insulator (HOTI) state at half-filling. An appropriate magnetic flux combined with NNN hopping can generate topological flat bands in the Chern insulator phase, exhibiting nontrivial Chern number and chiral edge states. The HOTI phase, in contrast, is characterized by topological corner states with quantized quadrupole moment within the bulk and edge states gap. We also present phase diagrams for the square-octagon lattice as functions of NNN and intercell hoppings under staggered magnetic fluxes. Our results indicate that the square-octagon lattice offers a promising platform for realizing topological flat bands, HOTI, and other topological and nontopological phases.

\end{abstract}

\maketitle

\section{Introduction}
Electronic states with energy dispersions $E_n(k)$ that are either linear or independent of momentum $k$ provide an emergent research frontier in condensed matter physics. A linear energy dispersion, $E_n(k) \propto k$, describes massless relativistic particles with high carrier velocity, as demonstrated in two-dimensional (2D) graphene and other nontrivial semimetals~\cite{novoselov2005two,PhysRevLett.95.226801,lee2008measurement,castro2009electronic}. In contrast, non-dispersive flat bands, characterized by constant energy dispersion, describe electrons with suppressed kinetic energy, leading to enhanced electron correlations~\cite{tang2011high,grushin2015characterization,balents2020superconductivity}. Such nondispersive flat bands arise naturally through Brillouin zone (BZ) folding in moir\'e superlattices or via destructive interference of electron wavefunctions on various decorated lattices~\cite{bistritzer2011moire,leykam2013flat,yan2014topological,leykam2018artificial,lim2020dirac,balents2020superconductivity,yin2022direct,liu2014exotic}. Electrons associated with flat bands are highly confined to specific lattice sites, resulting in strong electron interactions that are crucial for realizing many-body quantum phenomena~\cite{depenbrock2012nature,han2012fractionalized,liu2014exotic,jiang2019dichotomy}. 

\begin{figure}[t!]
\centering
\includegraphics[width=\columnwidth]{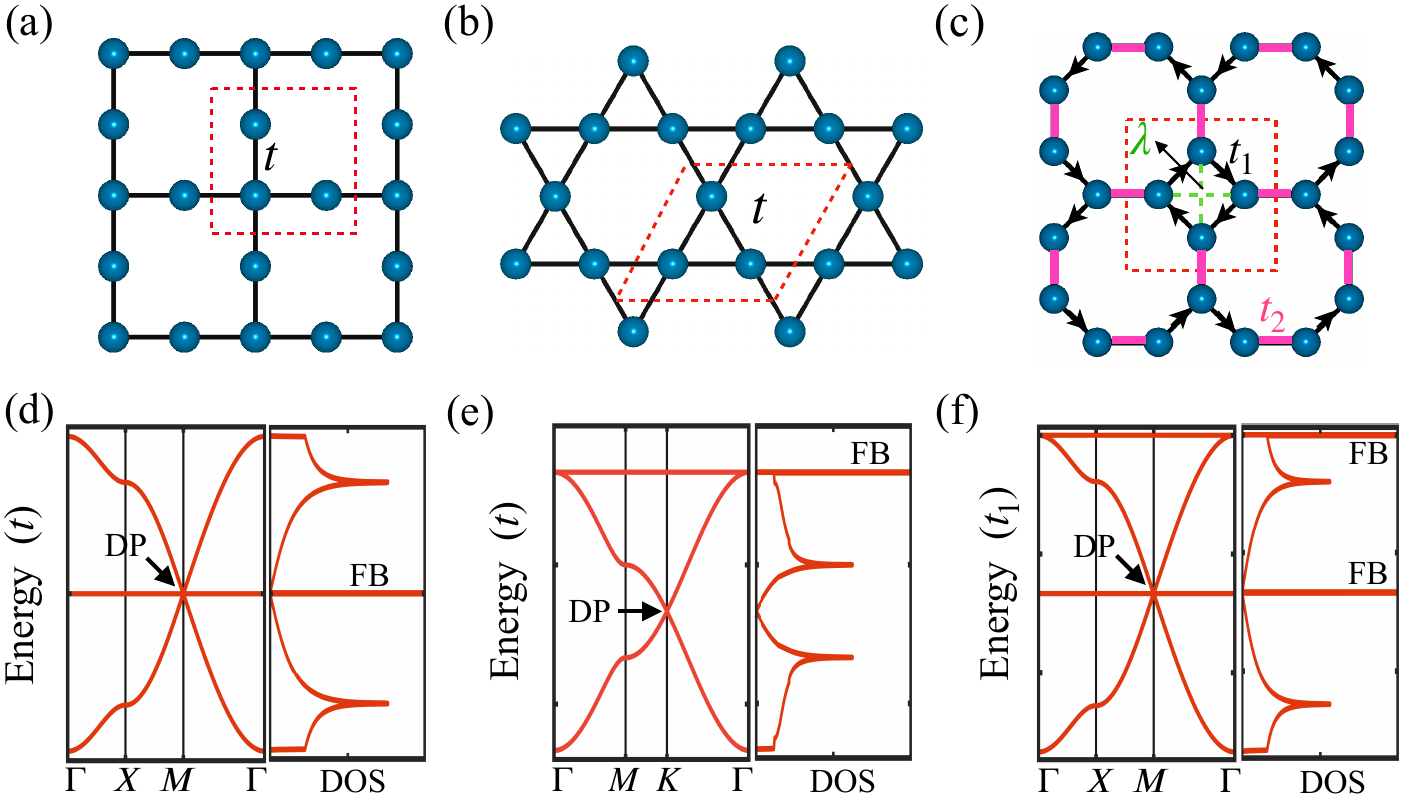}
\caption{Two-dimensional lattice models (top row) and their characteristic energy dispersions (bottom row) with Dirac points (DP) and flat bands (FB).  (a),(d) Lieb lattice and (b),(e) kagome lattice with three atomic sites per unit cell (dashed red lines) and nearest neighbor (NN) hopping $t$. In both lattices, the Dirac points are situated at the center of the energy spectrum. In the Lieb lattice, the flat bands lie at the center of the spectrum, while in the kagome lattice, the flat bands shift to the top of the energy spectrum.  The corresponding density of states reveals large diverging features at saddle points and flat band energies. (c),(f) Square octagon lattice with four atomic sites per unit cell. $t_1$, $\lambda$, and $t_2$ denote NN, next-nearest neighbor (NNN), and intercell hoppings, respectively. Arrows show the magnetic flux directions in the lattice (see text for details). The associated energy dispersion with two flat bands, one at the center and the other at the top of the spectrum is shown. } 
\label{fig1}
\end{figure}

Among the various two-dimensional lattices, the Lieb and kagome lattices are notable for displaying Dirac points, flat bands due to destructive wavefunction interference, and saddle-point van Hove singularities (VHSs)~\cite{mielke1991ferromagnetic,lieb1994flux,zhang2019kagome,graf2021designing}. In the Lieb lattice (Fig.~\ref{fig1}(a)), the flat bands are located at the center of the energy spectrum, with Dirac bands forming a three-fold degenerate point at the corner of the square BZ (Fig.~\ref{fig1}(d)). In contrast, in the kagome lattice (Fig.~\ref{fig1}(b)), the flat bands are found either at the top or bottom of the spectrum, while the Dirac nodes remain pinned at the center of the kagome bands (Fig.~\ref{fig1}(e)). External strain-induced interconversion between the Lieb and kagome lattices has revealed a transition of the flat bands from the middle of the energy spectrum to its extrema~\cite{jiang2019topological}.

The realization of a quantum anomalous Hall or Chern insulator state on a honeycomb lattice with staggered magnetic flux (Haldane model)~\cite{haldane1988model} has sparked global interest in the design of topological quantum states in a range of systems, including periodic lattices, amorphous materials, and quasicrystals~\cite{Singh2022,hasan2010,kane2005,he2019quasicrystalline,kraus2012}. Of particular interest is the interplay between nontrivial topology and flat bands, which leads to the emergence of nontrivial flat bands that provide an ideal platform for realizing correlated topological states. These topological flat bands can be engineered by systematically adjusting hopping potentials or phase factors in Chern insulator states~\cite{hu2011topological,sun2011nearly,wang2011nearly,guan2023staggered,he2024topological}. However, it is often observed that long-range hopping is essential for generating topological flat bands, making their experimental realization rare. Only a few systems have demonstrated topological flat bands within the Chern insulator phase by incorporating next-nearest-neighbor (NNN) hoppings with staggered magnetic fluxes~\cite{hu2011topological,sun2011nearly,wang2011nearly,guan2023staggered,he2024topological,he2022topological}. 

The Chern insulator phase is identified as a first-order topological insulator where $d$-dimensional topology manifests in ($d-1$)-dimensional boundary states, as outlined in bulk-boundary correspondence. In contrast, the higher-order topological insulators (HOTIs)~\cite{benalcazar2017electric,liu2017novel,schindler2018higher,franca2018anomalous,liu2019helical,guan2023staggered,he2022topological,he2024topological} are characterized by ($d-2$)-dimensional boundary states. For instance, a 2D HOTI exhibits gapped edge states with robust conducting corner states. Several studies have demonstrated that staggered magnetic flux or lattice dimerization can drive the topological phase transition between Chern insulator states and HOTIs~\cite{he2022topological,guan2023staggered,he2024topological}. Nevertheless, the tunability of topological flat bands across the energy spectrum using various internal or external parameters and the realization of HOTIs within the Chern insulator phase remains largely unexplored.

In this paper, we study the square-octagon lattice and show how controlling various unit cell hoppings and staggered magnetic fluxes can generate flat bands with nontrivial Chern numbers and HOTI states. The square-octagon lattice~\cite{kargarian2010topological,pal2018nontrivial,nunes2020flat,he2022topological1,yan2023intrinsic,he2023dirac} exhibits Dirac nodes, VHSs, and multiple flat bands (Figs.~\ref{fig1}(f)) and allows the possibility of tuning these states by adjusting intercell hoppings that control lattice site dimerization. At a low dimerization, a Chern insulator state emerges with a single band inversion and chiral edge modes as the magnetic flux changes. As dimerization increases, two band inversions occur, resulting in both edge and corner states, characteristic of the HOTI phase with a quantized quadrupole moment at half-filling. Additionally, we identify topological flat bands with a nontrivial Chern number ($C=\pm1$) and fully quenched kinetic energy. We also present phase diagrams based on nearest-neighbor (NN) and intercell hoppings, with and without staggered magnetic fields. Our findings demonstrate the tunability of topological flat bands and other topological states in the square-octagon lattice by controlling hopping potentials and magnetic fluxes.

\section{Tight-binding model}

We present the square-octagon lattice structure in Fig.~\ref{fig1}(c), highlighting various hoppings parameters and magnetic flux directions. The corresponding primitive unit cell (red dashed box) has four atomic sites. We define $t_1$, $t_2$, and $\lambda$ as the NN intracell hopping, intercell hopping, and NNN hopping, respectively. When $t_2 = t_1$, the lattice exhibits no lattice dimerization, whereas  $t_2 >> t_1$ represents a highly dimerized lattice.  A uniform magnetic flux applied perpendicular to the plane of each square plaquette induces an Aharonov-Bohm phase. As a result, the NN hoppings acquire a phase factor $e^{\pm i\theta}$, where $\theta=\frac{2\pi\Phi}{4\Phi_{0}}$, with $\Phi_0=\frac{hc}{e}$ represents the fundamental flux quantum. The $\pm$ sign denotes the direction of hoppings along the square loop, as indicated by arrows in Fig.~\ref{fig1}(c). The single-orbital spinless tight-binding Hamiltonian is defined as,

\begin{equation}
H = -t_{1}\sum_{\langle j,k \rangle}  e^{i\theta} c_j^\dag c_k -t_{2}\sum_{\langle j,k \rangle'} c_j^\dag c_k  -\lambda\sum_{\langle \langle j,k \rangle \rangle} c_j^\dag c_k + H.c.
\label{ham}
\end{equation}
Here, $c_j^\dag $ and  $c_j$ represent creation and annihilation operators for an electron at site $j$. The terms $\langle j,k \rangle $, $\langle j,k \rangle'$, and $ \langle \langle j,k \rangle \rangle$ correspond to NN, intercell, and NNN hoppings, respectively. 
The band structure of the square octagon lattice can be obtained by transforming Eq.~\ref{ham} into momentum space,
\begin{equation}
    H = \sum_{\vb*{k}} \psi^{\dag}_{\vb*{k}} \hat{\mathcal{H}}(\vb*{k}) \psi_{\vb*{k}}
\end{equation}
with  $\psi_{\vb*{k}} = (c_{1\vb*{k}}, c_{2\vb*{k}}, c_{3\vb*{k}}, c_{4\vb*{k}})$
and 
\begin{equation}
\small
\hat{\mathcal{H}}(\vb*{k}) = -\left[ \begin{array}{c c c c}
\epsilon & t_{1} e^{i\theta} & \lambda + t_{2} e^{-ik_{x}a} & t_{1}e^{-i\theta} \\
t_{1}e^{-i\theta} & \epsilon & t_{1}e^{i\theta} & \lambda + t_{2} e^{ik_{y}a} \\
\lambda+t_{2}e^{ik_{x}a} & t_{1} e^{-i\theta} & \epsilon & t_{1} e^{i\theta} \\
t_{1} e^{i\theta} & \lambda+t_{2}e^{-ik_{y}a} & t_{1} e^{-i\theta} & \epsilon
\end{array}
\right ] 
\label{hamk}
\end{equation}

When applying a staggered magnetic flux, the system breaks time-reversal symmetry,  potentially leading to the formation of Chern insulator states characterized by nontrivial Chern numbers. The Chern number associated with the $n^{th}$ band is determined by integrating the Berry curvature $\Omega_{n}(\vb*{k})$ over the full BZ~\cite{thouless1982quantized},
\begin{equation}
C_{n} = \frac{1}{2\pi}\int_{BZ}d^{2}\vb*{k}~\Omega_{n}(\vb*{k})
 \label{chern}
\end{equation}
The Berry curvature $\Omega_{n}(\vb*{k}) = \curl \mathbf{A}_{n}(\vb*{k})$, where $\mathbf{A}_{n}(\vb*{k})$ represents the Berry connection, $\mathbf{A}_{n}(\vb*{k}) = -i \langle u_{n\vb*{k}}|\grad_{\vb*{k}}|u_{n\vb*{k}} \rangle $ with $|u_{n\vb*{k}}\rangle$ denoting the Bloch state. The total Chern number of the system is expressed as $C=\sum_{n=1}^{N_{Occ}} C_{n}$, where $n$ is the band index.

The square-octagon lattice respects inversion symmetry $P\hat{\mathcal{H}}(\vb*{k})P^{\dag} = \hat{\mathcal{H}}(-\vb*{k})$ with the inversion operator 
\begin{equation}
P = -\left[ \begin{array}{c c c c}
0 & 0 & 0 & 1 \\
0 & 0 & 1 & 0 \\
0 & 1 & 0 & 0 \\
1 & 0 & 0 & 0
\end{array}
\right ] 
\label{parity}
\end{equation}
Inversion symmetry plays a crucial role in defining the bulk polarization $P_{i}$ along the $i^{th}$ direction and quadrupole moment $Q_{ij}$~\cite{fu2007topological,hughes2011inversion,fang2012bulk,liu2019helical}, which characterize the HOTI phase. The bulk polarization $P_{i}$ for the square-octagon lattice, expressed in terms of band parity  $\xi$, is given by 
\begin{equation}
P_{i}= \frac{1}{2} \left( \sum_{n}^{N_{\mathrm{Occ}}} q_{i}^{n}~ \mathrm{mod}~2 \right),~~~~~~~~ (-1)^{q_{i}^{n}} = \frac{\xi(X_{i})}{\xi(\Gamma)} 
\label{polariz}
\end{equation}
where, $i$ corresponds to either the $x-$ or $y-$direction, and $\xi(X_{i})$ denotes the band parity at high-symmetry point $X$ or $Y$. The quadrupole moment is related to the bulk polarization as
\begin{equation}
Q_{ij} = \sum_{n}^{N_{\mathrm{Occ}}}P_{i}^{n} P_{j}^{n}
\label{qudra}
\end{equation}
where $P_{i}^{n} = \frac{1}{2} (q_{i}^{n}~ \mathrm{mod}~2)$ represents the polarization associated with the $n^{th}$ band.

\begin{figure}[t!]
\centering
\includegraphics[width=\columnwidth]{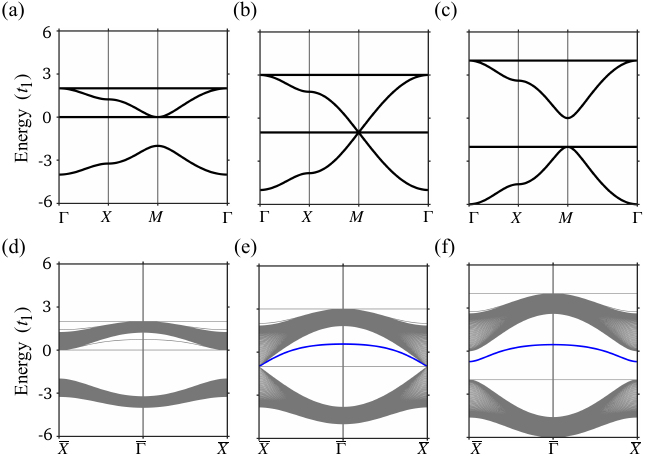}
\caption{Evolution of the bulk (top row) and edge band structure (bottom row) of the square-octagon lattice as a function of intercell hopping $t_2$. (a),(d) $t_2=t_1$, (b),(e) $t_2=2t_1$, and (c),(f) $t_2=3t_1$. The next-nearest neighbor (NNN) hopping is $\lambda=t_1$ with the intracell nearest-neighbor (NN) hopping set to $t_1=1$. Two non-dispersive flat bands at eigenvalues $E=t_2 \pm t_1$ appear in the spectrum. A topological phase transition from a trivial to a nontrivial state is evident. In (d)-(f), the projected bulk bands are shown in shaded grey, while the in-gap edge states are plotted in blue.}
\label{fig2}
\end{figure}

\section{Evolution of flat bands and Dirac states}\label{flatband}

We start discussing the evolution of flat bands and Dirac states in the square-octagon lattice in the absence of magnetic flux $\Phi=0$. The eigenvalues associated with Hamiltonian Eq.~\ref{hamk} after considering the NNN hopping $\lambda = t_1$ are
\begin{equation}
\left.
\begin{split}
E_{n=1-4} &= t_{1} \pm t_{2},\\ 
&-t_{1} \pm \sqrt{4 t_{1}^2 +t_{2}^2 + 2t_{1}t_{2} (cos{k_{x}}+cos{k_{y}})}
   \end{split}
   \right\}
   \label{sqoct_eigen}
\end{equation}
The band structure associated with Eq.~\ref{sqoct_eigen}, obtained along high-symmetry directions in the square BZ, is presented in Fig.~\ref{fig2}. This energy spectrum features two flat bands: one located at half-filling and another positioned at the top of the spectrum. These flat bands touch other bands to form Dirac cones at $M$ point for half-filling and $\Gamma$ point for $3/4$ filling. The flat bands are sensitive to variations in the intercell hopping $t_2$. When $t_2 < t_1$, the narrow bandwidth leads to a gap between the first band at $1/4$ filling and the second band at half-filling (Fig.~\ref{fig2}(a)). As $t_2$ increases gradually,  the bandwidth increases, resulting in a reduced gap between these two bands. The band gap closes when $t_2=2t_1$, transitioning the entire spectrum to a metallic state with the formation of a three-fold degenerate point at $M$ (Fig.~\ref{fig2}(b)). Beyond this point, with a further increase in $t_2$ (Fig.~\ref{fig2}(c)), a nontrivial gap emerges between the second and third bands at half-filling, indicating the formation of a topological insulator. The nontrivial nature of these states is confirmed by calculating the quantized bulk polarization using Eq.~\ref{polariz}. The edge state spectrum, illustrated in Figs.~\ref{fig2}(d)-(f), demonstrates the presence of nontrivial edge states within the band gap for $t_2 \ge 2t_1$. These results indicate a topological phase transition from a trivial metal to a nontrivial topological insulator, with an intermediate topological metallic phase arising from adjustments to the ratio of the intercell hopping  $t_2$ and NN hopping $t_1$. Notably, both the flat bands remain robust throughout this transition, though their energies can be tuned by varying the intercell hopping $t_2$.

\section{Chern insulator, HOTI, and topological flat bands}\label{tfb}

\begin{figure}[t!]
\centering
\includegraphics[width=\columnwidth]{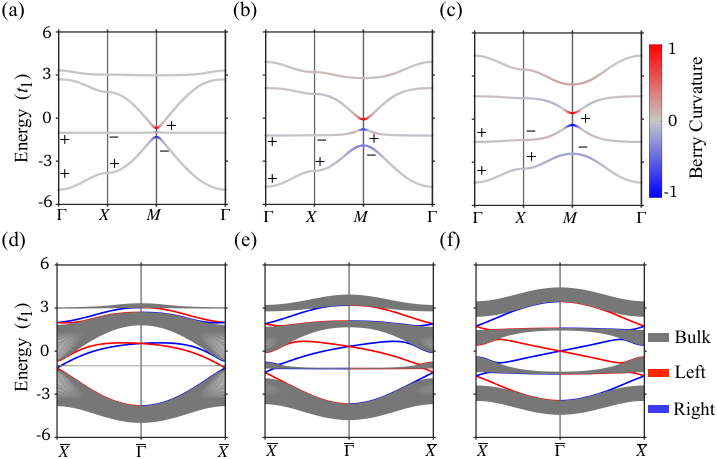}
\caption{Bulk (top row) and edge (bottom row) electronic states as a function of staggered magnetic flux. (a),(d) $\Phi=0.1\Phi_0$, (b),(e) $\Phi=0.3\Phi_0$, and (c),(f) $\Phi=0.5\Phi_0$. The hopping parameters used are $t_2=2t_1$, $\lambda=t_1$, and $t_1=1$,  corresponding to the band structure shown in Fig.~\ref{fig2}(b). In (a)-(c), band-resolved Berry curvature is shown with a color scale, and band parities are given with $\pm 1$. In (d)-(f), red and blue lines represent chiral edge states localized on opposite edges of the lattice, characterizing the Chern insulator phase.}
\label{fig3_soP}
\end{figure}

We now explore the emergence of various topological states in the presence of the staggered magnetic flux $\Phi$. Figure~\ref{fig3_soP} presents the energy spectrum of the square-octagon lattice, corresponding to Fig.~\ref{fig2}(b), but obtained for different values of staggered magnetic flux ($t_2=2t_1$, $\lambda=t_1$, and $t_1=1$). The application of magnetic flux breaks time-reversal symmetry, resulting in the opening of band gaps at the $M$ Dirac point for half-filling and $\Gamma$ point for $3/4$ filling. These gaps increase in size as the magnetic flux is enhanced. Importantly, both flat bands found in square-octagon lattice exhibit substantial dispersion under the influence of the magnetic flux. Despite their dispersion, the system maintains an insulating state across all filling levels, with the bands acquiring significant Berry curvature under $\Phi$, as illustrated in Figs.~\ref{fig3_soP}(a)-(c). The presence of band gaps allows the calculation of the Chern number using Eq.~\ref{chern}. We obtain nontrivial Chern numbers of  $C_n=-1$ and $+1$ for half-filling and $3/4$ filling, respectively. These results are consistent with the calculated parity of the bands at high-symmetry points. To substantiate the calculated nontrivial Chern numbers, we calculate the edge spectrum of the lattice under magnetic flux (Figs.~\ref{fig3_soP}(d)-(f)). The opposite moving chiral states on different edges are resolved between the projected bulk band gap. These results confirm that the system realizes a Chern insulator phase under the influence of magnetic flux. However, as the magnetic flux increases, the flat bands lose their characteristic flatness without acquiring any nontrivial Chern number. 

\begin{figure}[t!]
\centering
\includegraphics[width=\columnwidth]{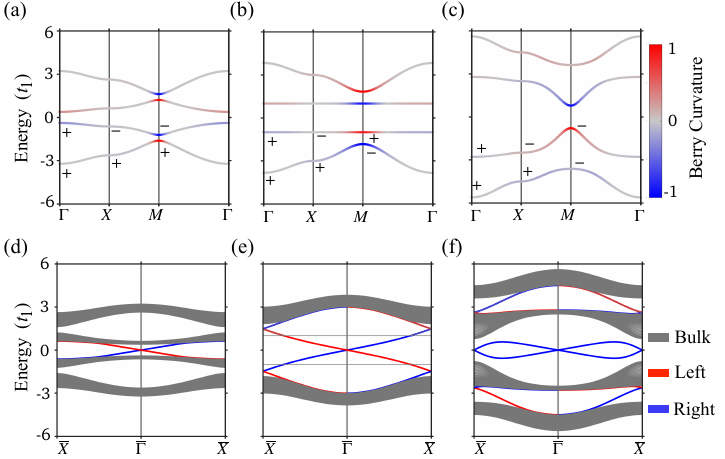}
\caption{Bulk (top row) and edge (bottom row) electronic states as a function of intercell hopping $t_2$ with staggered magnetic flux, $\Phi=0.5\Phi_0$. (a),(d) $t_2=0.8t_1$, (b),(e) $t_2=\frac{t_1}{\sqrt{2}}$, and (c),(f) $t_2=3.2t_1$. The NNN hopping $\lambda=t_1$ and intracell NN $t_1$ hopping is set to 1. Band-resolved Berry curvature is shown with a color scale and band parities are given with $\pm 1$ in (a)-(c). Red and blue lines in (d)-(f) represent chiral edge states localized on opposite edges of the lattice.}
\label{fig3}
\end{figure}

The calculated band structures for a fixed magnetic flux of $\Phi=0.5\Phi_0$, as a function of intercell hopping $t_2$, are presented in Fig.~\ref{fig3}. Introducing magnetic flux breaks time-reversal symmetry (see Fig.~\ref{fig3_soP}), opening gaps at the Dirac points. Interestingly, for a small value of  $t_2<0.8t_1$, the gaps at all the filling levels are topologically trivial. This is confirmed by both the calculation of the Chern number and the analysis of band parities at high-symmetry points. 
As $t_2$ increases, a band inversion occurs at the $\Gamma$ point between the second and third bands, resulting in an inverted band gap at half-filling (Fig.~\ref{fig3}(a)). The calculated Chern number associated with this state is $C=+1$, which gives rise to chiral edge states within the bulk gap (Fig.~\ref{fig3}(d)). With further increases in $t_2$, band inversions happen at $1/4$ and $3/4$ filling without affecting the nontrivial Chern number and edge states at half-filling.  When $t_2=3.2 t_{1}$, a second band inversion occurs at half-filling, resulting in a zero Chern number (Fig.~\ref{fig3}(c)). Nevertheless, the gapless states localized at the edges of the lattice remain within the bulk band gap, as illustrated in Fig.~\ref{fig3}(f). Importantly, even for $t_2>>t_1$, the edge states remain gapless, and the system continues to exhibit a zero Chern number.

\begin{figure*}
\includegraphics[width=2\columnwidth]{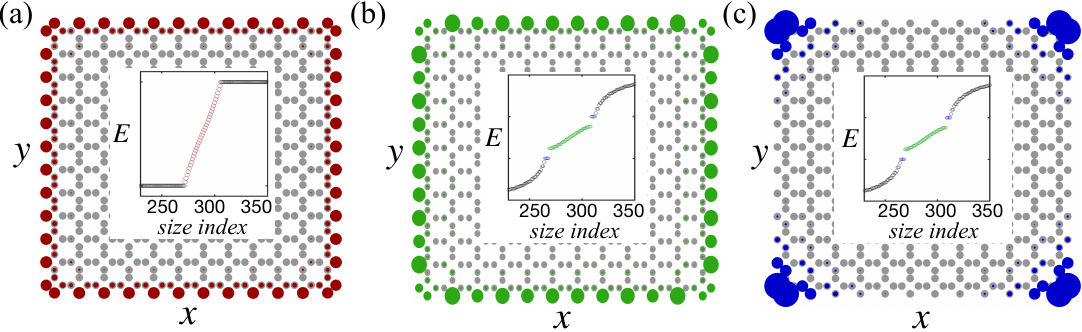}
\caption{Distribution of wavefunction $|\psi|^2$ for the Chern insulator and the HOTI state with staggered magnetic flux, $\Phi=0.5\Phi_0$. (a) Chern insulator state at $1/2$ filling with $t_2=\frac{t_1}{\sqrt{2}}$. (b) Gapless edge states and (c) corner state of the HOTI state with $t_{2}=3.2t_{1}$. The NNN hopping $\lambda=t_{1}$ and intracell NN $t_1$ hopping is fixed at 1. The bulk and the in-gap eigenstates are shown in the insets. Bulk eigen states are marked with black color and the edge states are marked with maroon color in (a) and green color in (b). The blue color denotes the corner states.} \label{higherorder}
\end{figure*}

To characterize the topological state associated with $t_2=3.2 t_{1}$ or at higher lattice dimerization ($t_2>>t_1$), we calculate the bulk polarization $P_{i=x,y}$ and quadrupole moment $Q_{xy}$ using Eqs.~\ref{polariz} and \ref{qudra}, respectively. The calculated results yield bulk polarizations of $(P_{x},P_{y})=(1/2,1/2)$ and a quadrupole moment $Q_{xy}=1/4$.  Such quantized nontrivial values of polarization and quadrupole moment suggest the realization of a HOTI state that can support corner states. Figure~\ref{higherorder} illustrates the real-space localization of states calculated using a nanodisk of the square-octagon lattice with an open boundary across various hopping parameters. At half-filling, the energy spectrum of the nanodisk remains gapless for $t_2=\frac{t_1}{\sqrt{2}}$, indicating a Chern insulator phase with uniformly distributed charge along the edges (see Fig.~\ref{higherorder}(a)). In contrast, for $t_2=3.2 t_{1}$, two distinct types of states emerge at half-filling: states localized at the edges (represented in green) and states localized at the corners (represented in blue), as shown in Figs.~\ref{higherorder}(b) and \ref{higherorder}(c). The corner states appear at energy eigenvalues of $E=\pm t_1$ within the gapped region and contribute to the nontrivial quadrupolar moment $Q_{xy}$ of $1/4$. It is important to note that a higher-order topological state in $d$ dimensions typically exhibits a gapped $d-1$ dimensional spectrum alongside a gapless $d-2$ spectrum. However, in the square-octagon lattice, we find that the gapless edge states remain robust even under high dimerization ($t_2>>t_1$), coexisting with gapless corner states, albeit at different energies. Such coexistence of gapless corner and edge states in square-octagon lattice makes it unique, providing a valuable platform for exploring dual topology under the same parameters.

\begin{figure}[ht]
\centering
\includegraphics[width=0.5\textwidth]{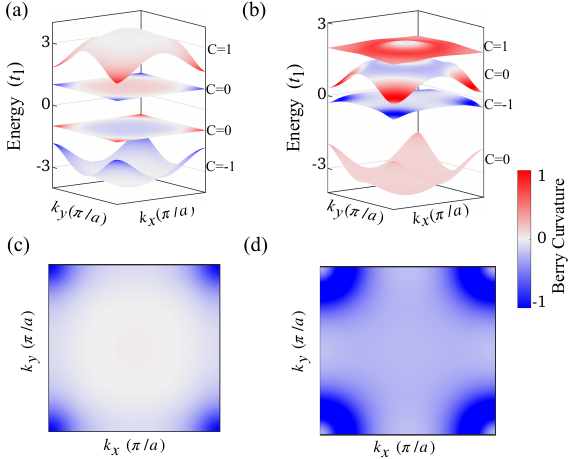}
\caption{Energy dispersion of bulk bands in the $k_x-k_y$ plane with staggered magnetic flux in Chern insulator phase for (a) $\Phi=0.5\Phi_{0}, t_{2}=t_{1}/cos(\theta)$ with $\theta=\pi/4, \lambda=t_{1}$ and (b) $\Phi=\Phi_{0}/10,t_{2}=t_{1},\lambda=t_{1}$. The Chern number ($C=\pm1$) for each band is marked.  A nearly flat band with nontrivial Chern number $C=-1$ appears at half-filling in (b). Panels (c)-(d) show the Berry curvature for the Chern bands at 1/4 filling in (a) and half-filling in (b). The color bar indicates the value of the Berry curvature, indicating the nontrivial nature of the bands.} 
\label{flat}
\end{figure}

The discussion above clearly indicates that the square-octagon lattice realizes various topological states both in the presence and absence of staggered magnetic flux. The characteristic flatness of the flat bands is removed under magnetic flux, and the bands gain substantial energy dispersion. However, we find that the flat bands can retain their dispersionless character under specific values of magnetic flux and hopping parameters. When the magnetic flux is set to $\Phi=n \Phi_{0}/2$, where $n$ is an odd integer and $\lambda=t_1$, this leads to theta $\theta=n\pi/4$ and generates a pair of flat bands at energies $E=\pm t_2 cos{\theta}$ for $t_2=t_1/cos{\theta}$ (Fig.~\ref{fig3}(b)). In Fig.~\ref{flat}(a), we present the calculated band structure within the full 2D BZ along with the calculated Chern number for each band when $\theta=\pi/4$. The first and fourth bands exhibit nontrivial Chern numbers of $-1$ and $+1$, respectively, while the flat bands display zero Chern numbers (Figs.~\ref{flat}(a),(c)). Notably, the Berry curvature associated with the flat bands at the zone corner is negative whereas it is positive at other parts of the BZ. This integrates to zero across the entire BZ. Nevertheless, the flat bands maintain nearly zero dispersion with an enhanced density of states in the electronic structure. 

Furthermore, when the magnetic flux is set to a small value of $\Phi=\Phi_{0}/10$ with $t_2=t_1$, the flat bands of the lattice transition to a nontrivial state, exhibiting a nonzero Chern number of  $C=\pm 1$ (Figs.~\ref{flat}(b),(d)). However, the flat bands gain a slight dispersion under these conditions, as shown in Fig.~\ref{flat}(b). The distribution of Berry curvature associated with a topological flat band at half-filling is presented in Fig.~\ref{flat}(d). The Berry curvature is highly localized at the corner of the BZ, near the Dirac points where band inversion occurs. 
These results demonstrate that both trivial and nontrivial flat bands can be realized in a square-octagon lattice under specifically tuned magnetic flux and hoping parameters.

\begin{figure}[ht]
\centering
\includegraphics[width=0.5\textwidth]{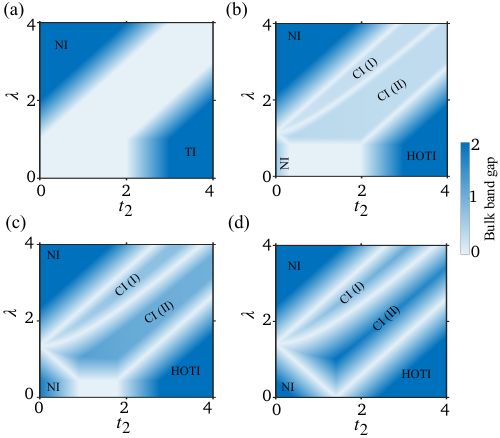}
\caption{Bandgap (in colorscale) of the square-octagon lattice as a function of intercell hopping $t_{2}$ and next-nearest-neighbor (NNN) hopping $\lambda$ for varying magnetic flux values. (a) $\Phi=0$, (b) $\Phi=0.1 \Phi_0$, (c) $\Phi=0.3 \Phi_0$, and (d) $\Phi=0.5 \Phi_0$. Bandgap is described between the second and third bands at half-filling. Topological states are indicated: NI (normal insulator), TI (topological insulator), CI (Chern insulator), and HOTI (higher-order topological insulator). CI(I) and CI(II) mark two Chern insulator phases (see text for details).} 
\label{phasehop}
\end{figure}

\section{Topological phase diagram}\label{phase}

We now present the calculated topological phases of square-octagon lattice by varying the intercell hopping $t_2$ and NNN hopping $\lambda$ parameters in Fig.~\ref{phasehop}.  The bandgap is defined between the second and third bands under half-filling conditions. In the absence of magnetic flux (Fig.~\ref{phasehop}(a)), we identify three distinct phases, normal insulator, topological metal, and topological insulator. Critical threshold values of $t_2$ and $\lambda$ are necessary to transition the system away from, a topological metal state, characterized by a three-fold Dirac cone at the $M$-point (see Fig.~\ref{fig2}(a)). Upon the introduction of magnetic flux, the energy spectrum becomes gaped, with a local gap at all the band crossings, leading to the emergence of Chern insulator and HOTI states. The phase transition between various topological states is accompanied by the closure of the bandgap at the half-filling, forming a critical point metallic state.  For a small magnetic flux of $\Phi=0.1\Phi_0$ (Fig.~\ref{phasehop}(b)), with small intercell hoping $t_2$, the system exhibits a Chern insulator phase at low NNN hoping $\lambda$. In contrast, at larger values of $t_2$, a HOTI phase with quantized quadrupolar charge and localized corner modes is realized. As the magnetic flux increases, the HOTI phase appears at the relatively smaller values of $t_2$, as shown in Figs.~\ref{phasehop}(c)-(d). Importantly, the HOTI state emerges only beyond specific thresholds of magnetic flux and $t_2$, while the Chern insulator state is formed at lower magnetic flux values.  We observe two distinct Chern insulator states, labelled CI(I) and CI(II) in Fig.~\ref{phasehop}. In the first Chern insulator phase (CI(I)), chiral edge states are present exclusively at half-filling. In contrast, the second Chern insulator phase (CI(II)) supports chiral edge states across all band fillings. Thus, the square-octagon lattice manifests a variety of topological phases with associated nontrivial edge or corner modes.\\

\section{Summary and conclusions}

We have investigated the square-octagon lattice to explore the emergence of topological flat bands and various nontrivial phases by tuning unit cell hopping parameters and staggered magnetic fluxes. This lattice exhibits characteristics similar to the Lieb and kagome lattices, featuring Dirac cones, flat bands, and VHSs in its electronic spectrum, with one flat band located at the center and another at the top of the spectrum. Our analysis shows that by adjusting the intercell hopping parameter $t_2$ (lattice dimerization), flat bands can be tuned across the Fermi level, facilitating a phase transition into a topological state with nontrivial edge states. Upon applying staggered magnetic flux, the system transitions into a Chern insulator phase at low lattice dimerization and into a HOTI phase at higher dimerization. The Chern insulator phase exhibits chiral edge states along both truncated edges, while the HOTI phase is characterized by a nontrivial quadrupolar moment and localized corner states. Importantly, we find that the higher-order topological phase features coexisting metallic edge modes and corner states, though their energy locations are different. We also show that a suitable combination of staggered magnetic flux and various hopping integrals can generate a topological flat band with a Chern number $C=\pm 1$. We present topological phase diagrams by tuning NN and intercell hoppings in the presence and absence of staggered magnetic fluxes. Our results demonstrate that the square-octagon lattice serves as a viable model for realizing tunable topological phases and flat bands through the control of hopping potentials and magnetic fluxes.

\section*{Acknowledgement}
We thank R. Verma and B. Patra for helpful discussions. This work is supported by the Department of Atomic Energy of the Government of India and benefited from the computational resources of TIFR, Mumbai. 

\bibliography{ref}

\begin{thebibliography}{49}%
\makeatletter
\providecommand \@ifxundefined [1]{%
 \@ifx{#1\undefined}
}%
\providecommand \@ifnum [1]{%
 \ifnum #1\expandafter \@firstoftwo
 \else \expandafter \@secondoftwo
 \fi
}%
\providecommand \@ifx [1]{%
 \ifx #1\expandafter \@firstoftwo
 \else \expandafter \@secondoftwo
 \fi
}%
\providecommand \natexlab [1]{#1}%
\providecommand \enquote  [1]{``#1''}%
\providecommand \bibnamefont  [1]{#1}%
\providecommand \bibfnamefont [1]{#1}%
\providecommand \citenamefont [1]{#1}%
\providecommand \href@noop [0]{\@secondoftwo}%
\providecommand \href [0]{\begingroup \@sanitize@url \@href}%
\providecommand \@href[1]{\@@startlink{#1}\@@href}%
\providecommand \@@href[1]{\endgroup#1\@@endlink}%
\providecommand \@sanitize@url [0]{\catcode `\\12\catcode `\$12\catcode
  `\&12\catcode `\#12\catcode `\^12\catcode `\_12\catcode `\%12\relax}%
\providecommand \@@startlink[1]{}%
\providecommand \@@endlink[0]{}%
\providecommand \url  [0]{\begingroup\@sanitize@url \@url }%
\providecommand \@url [1]{\endgroup\@href {#1}{\urlprefix }}%
\providecommand \urlprefix  [0]{URL }%
\providecommand \Eprint [0]{\href }%
\providecommand \doibase [0]{https://doi.org/}%
\providecommand \selectlanguage [0]{\@gobble}%
\providecommand \bibinfo  [0]{\@secondoftwo}%
\providecommand \bibfield  [0]{\@secondoftwo}%
\providecommand \translation [1]{[#1]}%
\providecommand \BibitemOpen [0]{}%
\providecommand \bibitemStop [0]{}%
\providecommand \bibitemNoStop [0]{.\EOS\space}%
\providecommand \EOS [0]{\spacefactor3000\relax}%
\providecommand \BibitemShut  [1]{\csname bibitem#1\endcsname}%
\let\auto@bib@innerbib\@empty
\bibitem [{\citenamefont {Novoselov}\ \emph {et~al.}(2005)\citenamefont
  {Novoselov}, \citenamefont {Geim}, \citenamefont {Morozov}, \citenamefont
  {Jiang}, \citenamefont {Katsnelson}, \citenamefont {Grigorieva},
  \citenamefont {Dubonos},\ and\ \citenamefont {Firsov}}]{novoselov2005two}%
  \BibitemOpen
  \bibfield  {author} {\bibinfo {author} {\bibfnamefont {K.~S.}\ \bibnamefont
  {Novoselov}}, \bibinfo {author} {\bibfnamefont {A.~K.}\ \bibnamefont {Geim}},
  \bibinfo {author} {\bibfnamefont {S.~V.}\ \bibnamefont {Morozov}}, \bibinfo
  {author} {\bibfnamefont {D.}~\bibnamefont {Jiang}}, \bibinfo {author}
  {\bibfnamefont {M.~I.}\ \bibnamefont {Katsnelson}}, \bibinfo {author}
  {\bibfnamefont {I.~V.}\ \bibnamefont {Grigorieva}}, \bibinfo {author}
  {\bibfnamefont {S.~V.}\ \bibnamefont {Dubonos}},\ and\ \bibinfo {author}
  {\bibfnamefont {A.~A.}\ \bibnamefont {Firsov}},\ }\bibfield  {title}
  {\bibinfo {title} {Two-dimensional gas of massless dirac fermions in
  graphene},\ }\href {https://doi.org/10.1038/nature04233} {\bibfield
  {journal} {\bibinfo  {journal} {Nature}\ }\textbf {\bibinfo {volume} {438}},\
  \bibinfo {pages} {197} (\bibinfo {year} {2005})}\BibitemShut {NoStop}%
\bibitem [{\citenamefont {Kane}\ and\ \citenamefont
  {Mele}(2005{\natexlab{a}})}]{PhysRevLett.95.226801}%
  \BibitemOpen
  \bibfield  {author} {\bibinfo {author} {\bibfnamefont {C.~L.}\ \bibnamefont
  {Kane}}\ and\ \bibinfo {author} {\bibfnamefont {E.~J.}\ \bibnamefont
  {Mele}},\ }\bibfield  {title} {\bibinfo {title} {Quantum spin hall effect in
  graphene},\ }\href {https://doi.org/10.1103/PhysRevLett.95.226801} {\bibfield
   {journal} {\bibinfo  {journal} {Phys. Rev. Lett.}\ }\textbf {\bibinfo
  {volume} {95}},\ \bibinfo {pages} {226801} (\bibinfo {year}
  {2005}{\natexlab{a}})}\BibitemShut {NoStop}%
\bibitem [{\citenamefont {Lee}\ \emph {et~al.}(2008)\citenamefont {Lee},
  \citenamefont {Wei}, \citenamefont {Kysar},\ and\ \citenamefont
  {Hone}}]{lee2008measurement}%
  \BibitemOpen
  \bibfield  {author} {\bibinfo {author} {\bibfnamefont {C.}~\bibnamefont
  {Lee}}, \bibinfo {author} {\bibfnamefont {X.}~\bibnamefont {Wei}}, \bibinfo
  {author} {\bibfnamefont {J.~W.}\ \bibnamefont {Kysar}},\ and\ \bibinfo
  {author} {\bibfnamefont {J.}~\bibnamefont {Hone}},\ }\bibfield  {title}
  {\bibinfo {title} {Measurement of the elastic properties and intrinsic
  strength of monolayer graphene},\ }\href
  {https://doi.org/10.1126/science.1157996} {\bibfield  {journal} {\bibinfo
  {journal} {Science}\ }\textbf {\bibinfo {volume} {321}},\ \bibinfo {pages}
  {385} (\bibinfo {year} {2008})}\BibitemShut {NoStop}%
\bibitem [{\citenamefont {Castro~Neto}\ \emph {et~al.}(2009)\citenamefont
  {Castro~Neto}, \citenamefont {Guinea}, \citenamefont {Peres}, \citenamefont
  {Novoselov},\ and\ \citenamefont {Geim}}]{castro2009electronic}%
  \BibitemOpen
  \bibfield  {author} {\bibinfo {author} {\bibfnamefont {A.~H.}\ \bibnamefont
  {Castro~Neto}}, \bibinfo {author} {\bibfnamefont {F.}~\bibnamefont {Guinea}},
  \bibinfo {author} {\bibfnamefont {N.~M.}\ \bibnamefont {Peres}}, \bibinfo
  {author} {\bibfnamefont {K.~S.}\ \bibnamefont {Novoselov}},\ and\ \bibinfo
  {author} {\bibfnamefont {A.~K.}\ \bibnamefont {Geim}},\ }\bibfield  {title}
  {\bibinfo {title} {The electronic properties of graphene},\ }\href
  {https://doi.org/10.1103/RevModPhys.81.109} {\bibfield  {journal} {\bibinfo
  {journal} {Rev. Mod. Phys.}\ }\textbf {\bibinfo {volume} {81}},\ \bibinfo
  {pages} {109} (\bibinfo {year} {2009})}\BibitemShut {NoStop}%
\bibitem [{\citenamefont {Tang}\ \emph {et~al.}(2011)\citenamefont {Tang},
  \citenamefont {Mei},\ and\ \citenamefont {Wen}}]{tang2011high}%
  \BibitemOpen
  \bibfield  {author} {\bibinfo {author} {\bibfnamefont {E.}~\bibnamefont
  {Tang}}, \bibinfo {author} {\bibfnamefont {J.-W.}\ \bibnamefont {Mei}},\ and\
  \bibinfo {author} {\bibfnamefont {X.-G.}\ \bibnamefont {Wen}},\ }\bibfield
  {title} {\bibinfo {title} {High-temperature fractional quantum hall states},\
  }\href {https://doi.org/10.1103/PhysRevLett.106.236802} {\bibfield  {journal}
  {\bibinfo  {journal} {Phys. Rev. Lett.}\ }\textbf {\bibinfo {volume} {106}},\
  \bibinfo {pages} {236802} (\bibinfo {year} {2011})}\BibitemShut {NoStop}%
\bibitem [{\citenamefont {Grushin}\ \emph {et~al.}(2015)\citenamefont
  {Grushin}, \citenamefont {Motruk}, \citenamefont {Zaletel},\ and\
  \citenamefont {Pollmann}}]{grushin2015characterization}%
  \BibitemOpen
  \bibfield  {author} {\bibinfo {author} {\bibfnamefont {A.~G.}\ \bibnamefont
  {Grushin}}, \bibinfo {author} {\bibfnamefont {J.}~\bibnamefont {Motruk}},
  \bibinfo {author} {\bibfnamefont {M.~P.}\ \bibnamefont {Zaletel}},\ and\
  \bibinfo {author} {\bibfnamefont {F.}~\bibnamefont {Pollmann}},\ }\bibfield
  {title} {\bibinfo {title} {Characterization and stability of a fermionic
  $\nu$= 1/3 fractional chern insulator},\ }\href
  {https://doi.org/10.1103/PhysRevB.91.035136} {\bibfield  {journal} {\bibinfo
  {journal} {Phys. Rev. B}\ }\textbf {\bibinfo {volume} {91}},\ \bibinfo
  {pages} {035136} (\bibinfo {year} {2015})}\BibitemShut {NoStop}%
\bibitem [{\citenamefont {Balents}\ \emph {et~al.}(2020)\citenamefont
  {Balents}, \citenamefont {Dean}, \citenamefont {Efetov},\ and\ \citenamefont
  {Young}}]{balents2020superconductivity}%
  \BibitemOpen
  \bibfield  {author} {\bibinfo {author} {\bibfnamefont {L.}~\bibnamefont
  {Balents}}, \bibinfo {author} {\bibfnamefont {C.~R.}\ \bibnamefont {Dean}},
  \bibinfo {author} {\bibfnamefont {D.~K.}\ \bibnamefont {Efetov}},\ and\
  \bibinfo {author} {\bibfnamefont {A.~F.}\ \bibnamefont {Young}},\ }\bibfield
  {title} {\bibinfo {title} {Superconductivity and strong correlations in
  moir{\'e} flat bands},\ }\href
  {https://doi.org/https://doi.org/10.1038/s41567-020-0906-9} {\bibfield
  {journal} {\bibinfo  {journal} {Nat. Phys.}\ }\textbf {\bibinfo {volume}
  {16}},\ \bibinfo {pages} {725} (\bibinfo {year} {2020})}\BibitemShut
  {NoStop}%
\bibitem [{\citenamefont {Bistritzer}\ and\ \citenamefont
  {MacDonald}(2011)}]{bistritzer2011moire}%
  \BibitemOpen
  \bibfield  {author} {\bibinfo {author} {\bibfnamefont {R.}~\bibnamefont
  {Bistritzer}}\ and\ \bibinfo {author} {\bibfnamefont {A.~H.}\ \bibnamefont
  {MacDonald}},\ }\bibfield  {title} {\bibinfo {title} {Moir{\'e} bands in
  twisted double-layer graphene},\ }\href@noop {} {\bibfield  {journal}
  {\bibinfo  {journal} {Proceedings of the National Academy of Sciences}\
  }\textbf {\bibinfo {volume} {108}},\ \bibinfo {pages} {12233} (\bibinfo
  {year} {2011})}\BibitemShut {NoStop}%
\bibitem [{\citenamefont {Leykam}\ \emph {et~al.}(2013)\citenamefont {Leykam},
  \citenamefont {Flach}, \citenamefont {Bahat-Treidel},\ and\ \citenamefont
  {Desyatnikov}}]{leykam2013flat}%
  \BibitemOpen
  \bibfield  {author} {\bibinfo {author} {\bibfnamefont {D.}~\bibnamefont
  {Leykam}}, \bibinfo {author} {\bibfnamefont {S.}~\bibnamefont {Flach}},
  \bibinfo {author} {\bibfnamefont {O.}~\bibnamefont {Bahat-Treidel}},\ and\
  \bibinfo {author} {\bibfnamefont {A.~S.}\ \bibnamefont {Desyatnikov}},\
  }\bibfield  {title} {\bibinfo {title} {Flat band states: Disorder and
  nonlinearity},\ }\href {https://doi.org/10.1103/PhysRevB.88.224203}
  {\bibfield  {journal} {\bibinfo  {journal} {Phys. Rev. B}\ }\textbf {\bibinfo
  {volume} {88}},\ \bibinfo {pages} {224203} (\bibinfo {year}
  {2013})}\BibitemShut {NoStop}%
\bibitem [{\citenamefont {Yan}\ and\ \citenamefont
  {Wan}(2014)}]{yan2014topological}%
  \BibitemOpen
  \bibfield  {author} {\bibinfo {author} {\bibfnamefont {Z.}~\bibnamefont
  {Yan}}\ and\ \bibinfo {author} {\bibfnamefont {S.}~\bibnamefont {Wan}},\
  }\bibfield  {title} {\bibinfo {title} {Topological phases, topological flat
  bands, and topological excitations in a one-dimensional dimerized lattice
  with spin-orbit coupling},\ }\href
  {https://doi.org/10.1209/0295-5075/107/47007} {\bibfield  {journal} {\bibinfo
   {journal} {Europhys. Lett.}\ }\textbf {\bibinfo {volume} {107}},\ \bibinfo
  {pages} {47007} (\bibinfo {year} {2014})}\BibitemShut {NoStop}%
\bibitem [{\citenamefont {Leykam}\ \emph {et~al.}(2018)\citenamefont {Leykam},
  \citenamefont {Andreanov},\ and\ \citenamefont
  {Flach}}]{leykam2018artificial}%
  \BibitemOpen
  \bibfield  {author} {\bibinfo {author} {\bibfnamefont {D.}~\bibnamefont
  {Leykam}}, \bibinfo {author} {\bibfnamefont {A.}~\bibnamefont {Andreanov}},\
  and\ \bibinfo {author} {\bibfnamefont {S.}~\bibnamefont {Flach}},\ }\bibfield
   {title} {\bibinfo {title} {Artificial flat band systems: from lattice models
  to experiments},\ }\href {https://doi.org/10.1080/23746149.2018.1473052}
  {\bibfield  {journal} {\bibinfo  {journal} {Adv. Phys.: X}\ }\textbf
  {\bibinfo {volume} {3}},\ \bibinfo {pages} {1473052} (\bibinfo {year}
  {2018})}\BibitemShut {NoStop}%
\bibitem [{\citenamefont {Lim}\ \emph {et~al.}(2020)\citenamefont {Lim},
  \citenamefont {Fuchs}, \citenamefont {Pi{\'e}chon},\ and\ \citenamefont
  {Montambaux}}]{lim2020dirac}%
  \BibitemOpen
  \bibfield  {author} {\bibinfo {author} {\bibfnamefont {L.-K.}\ \bibnamefont
  {Lim}}, \bibinfo {author} {\bibfnamefont {J.-N.}\ \bibnamefont {Fuchs}},
  \bibinfo {author} {\bibfnamefont {F.}~\bibnamefont {Pi{\'e}chon}},\ and\
  \bibinfo {author} {\bibfnamefont {G.}~\bibnamefont {Montambaux}},\ }\bibfield
   {title} {\bibinfo {title} {Dirac points emerging from flat bands in
  lieb-kagome lattices},\ }\href {https://doi.org/10.1103/PhysRevB.101.045131}
  {\bibfield  {journal} {\bibinfo  {journal} {Phys. Rev. B}\ }\textbf {\bibinfo
  {volume} {101}},\ \bibinfo {pages} {045131} (\bibinfo {year}
  {2020})}\BibitemShut {NoStop}%
\bibitem [{\citenamefont {Yin}\ \emph {et~al.}(2022)\citenamefont {Yin},
  \citenamefont {Tong}, \citenamefont {Zhou}, \citenamefont {Zhang},
  \citenamefont {Tian}, \citenamefont {Zhang}, \citenamefont {Zhang},\ and\
  \citenamefont {Qin}}]{yin2022direct}%
  \BibitemOpen
  \bibfield  {author} {\bibinfo {author} {\bibfnamefont {L.-J.}\ \bibnamefont
  {Yin}}, \bibinfo {author} {\bibfnamefont {L.-H.}\ \bibnamefont {Tong}},
  \bibinfo {author} {\bibfnamefont {Y.-Y.}\ \bibnamefont {Zhou}}, \bibinfo
  {author} {\bibfnamefont {Y.}~\bibnamefont {Zhang}}, \bibinfo {author}
  {\bibfnamefont {Y.}~\bibnamefont {Tian}}, \bibinfo {author} {\bibfnamefont
  {L.}~\bibnamefont {Zhang}}, \bibinfo {author} {\bibfnamefont
  {L.}~\bibnamefont {Zhang}},\ and\ \bibinfo {author} {\bibfnamefont
  {Z.}~\bibnamefont {Qin}},\ }\bibfield  {title} {\bibinfo {title} {Direct
  observation of moir{\'e} flat-band breakdown at the edge of magic-angle
  twisted bilayer graphene},\ }\href
  {https://doi.org/https://doi.org/10.1103/PhysRevB.105.L201405} {\bibfield
  {journal} {\bibinfo  {journal} {Phys. Rev. B}\ }\textbf {\bibinfo {volume}
  {105}},\ \bibinfo {pages} {L201405} (\bibinfo {year} {2022})}\BibitemShut
  {NoStop}%
\bibitem [{\citenamefont {Liu}\ \emph {et~al.}(2014)\citenamefont {Liu},
  \citenamefont {Liu},\ and\ \citenamefont {Wu}}]{liu2014exotic}%
  \BibitemOpen
  \bibfield  {author} {\bibinfo {author} {\bibfnamefont {Z.}~\bibnamefont
  {Liu}}, \bibinfo {author} {\bibfnamefont {F.}~\bibnamefont {Liu}},\ and\
  \bibinfo {author} {\bibfnamefont {Y.-S.}\ \bibnamefont {Wu}},\ }\bibfield
  {title} {\bibinfo {title} {Exotic electronic states in the world of flat
  bands: From theory to material},\ }\href
  {https://doi.org/10.1088/1674-1056/23/7/077308} {\bibfield  {journal}
  {\bibinfo  {journal} {Chin. Phys. B}\ }\textbf {\bibinfo {volume} {23}},\
  \bibinfo {pages} {077308} (\bibinfo {year} {2014})}\BibitemShut {NoStop}%
\bibitem [{\citenamefont {Depenbrock}\ \emph {et~al.}(2012)\citenamefont
  {Depenbrock}, \citenamefont {McCulloch},\ and\ \citenamefont
  {Schollw{\"o}ck}}]{depenbrock2012nature}%
  \BibitemOpen
  \bibfield  {author} {\bibinfo {author} {\bibfnamefont {S.}~\bibnamefont
  {Depenbrock}}, \bibinfo {author} {\bibfnamefont {I.~P.}\ \bibnamefont
  {McCulloch}},\ and\ \bibinfo {author} {\bibfnamefont {U.}~\bibnamefont
  {Schollw{\"o}ck}},\ }\bibfield  {title} {\bibinfo {title} {Nature of the
  spin-liquid ground state of the s= 1/2 heisenberg model on the kagome
  lattice},\ }\href {https://doi.org/10.48550/arXiv.1205.4858} {\bibfield
  {journal} {\bibinfo  {journal} {Phys. Rev. lett.}\ }\textbf {\bibinfo
  {volume} {109}},\ \bibinfo {pages} {067201} (\bibinfo {year}
  {2012})}\BibitemShut {NoStop}%
\bibitem [{\citenamefont {Han}\ \emph {et~al.}(2012)\citenamefont {Han},
  \citenamefont {Helton}, \citenamefont {Chu}, \citenamefont {Nocera},
  \citenamefont {Rodriguez-Rivera}, \citenamefont {Broholm},\ and\
  \citenamefont {Lee}}]{han2012fractionalized}%
  \BibitemOpen
  \bibfield  {author} {\bibinfo {author} {\bibfnamefont {T.-H.}\ \bibnamefont
  {Han}}, \bibinfo {author} {\bibfnamefont {J.~S.}\ \bibnamefont {Helton}},
  \bibinfo {author} {\bibfnamefont {S.}~\bibnamefont {Chu}}, \bibinfo {author}
  {\bibfnamefont {D.~G.}\ \bibnamefont {Nocera}}, \bibinfo {author}
  {\bibfnamefont {J.~A.}\ \bibnamefont {Rodriguez-Rivera}}, \bibinfo {author}
  {\bibfnamefont {C.}~\bibnamefont {Broholm}},\ and\ \bibinfo {author}
  {\bibfnamefont {Y.~S.}\ \bibnamefont {Lee}},\ }\bibfield  {title} {\bibinfo
  {title} {Fractionalized excitations in the spin-liquid state of a
  kagome-lattice antiferromagnet},\ }\href
  {https://doi.org/10.1038/nature11659} {\bibfield  {journal} {\bibinfo
  {journal} {Nature}\ }\textbf {\bibinfo {volume} {492}},\ \bibinfo {pages}
  {406} (\bibinfo {year} {2012})}\BibitemShut {NoStop}%
\bibitem [{\citenamefont {Jiang}\ \emph
  {et~al.}(2019{\natexlab{a}})\citenamefont {Jiang}, \citenamefont {Liu},
  \citenamefont {Mei}, \citenamefont {Cui},\ and\ \citenamefont
  {Liu}}]{jiang2019dichotomy}%
  \BibitemOpen
  \bibfield  {author} {\bibinfo {author} {\bibfnamefont {W.}~\bibnamefont
  {Jiang}}, \bibinfo {author} {\bibfnamefont {Z.}~\bibnamefont {Liu}}, \bibinfo
  {author} {\bibfnamefont {J.-W.}\ \bibnamefont {Mei}}, \bibinfo {author}
  {\bibfnamefont {B.}~\bibnamefont {Cui}},\ and\ \bibinfo {author}
  {\bibfnamefont {F.}~\bibnamefont {Liu}},\ }\bibfield  {title} {\bibinfo
  {title} {Dichotomy between frustrated local spins and conjugated electrons in
  a two-dimensional metal--organic framework},\ }\href
  {https://doi.org/10.1039/C8NR08479C} {\bibfield  {journal} {\bibinfo
  {journal} {Nanoscale}\ }\textbf {\bibinfo {volume} {11}},\ \bibinfo {pages}
  {955} (\bibinfo {year} {2019}{\natexlab{a}})}\BibitemShut {NoStop}%
\bibitem [{\citenamefont {Mielke}(1991)}]{mielke1991ferromagnetic}%
  \BibitemOpen
  \bibfield  {author} {\bibinfo {author} {\bibfnamefont {A.}~\bibnamefont
  {Mielke}},\ }\bibfield  {title} {\bibinfo {title} {Ferromagnetic ground
  states for the hubbard model on line graphs},\ }\href
  {https://doi.org/10.1088/0305-4470/24/2/005} {\bibfield  {journal} {\bibinfo
  {journal} {J. Phys. A: Mathematical and General}\ }\textbf {\bibinfo {volume}
  {24}},\ \bibinfo {pages} {L73} (\bibinfo {year} {1991})}\BibitemShut
  {NoStop}%
\bibitem [{\citenamefont {Lieb}(1994)}]{lieb1994flux}%
  \BibitemOpen
  \bibfield  {author} {\bibinfo {author} {\bibfnamefont {E.~H.}\ \bibnamefont
  {Lieb}},\ }\bibfield  {title} {\bibinfo {title} {Flux phase of the
  half-filled band},\ }\href {https://doi.org/10.1103/PhysRevLett.73.2158}
  {\bibfield  {journal} {\bibinfo  {journal} {Phys. Rev. Lett.}\ }\textbf
  {\bibinfo {volume} {73}},\ \bibinfo {pages} {2158} (\bibinfo {year}
  {1994})}\BibitemShut {NoStop}%
\bibitem [{\citenamefont {Zhang}\ \emph {et~al.}(2019)\citenamefont {Zhang},
  \citenamefont {Kang}, \citenamefont {Huang}, \citenamefont {Jiang},
  \citenamefont {Ni}, \citenamefont {Kang}, \citenamefont {Zhang},
  \citenamefont {Xu}, \citenamefont {Liu},\ and\ \citenamefont
  {Liu}}]{zhang2019kagome}%
  \BibitemOpen
  \bibfield  {author} {\bibinfo {author} {\bibfnamefont {S.}~\bibnamefont
  {Zhang}}, \bibinfo {author} {\bibfnamefont {M.}~\bibnamefont {Kang}},
  \bibinfo {author} {\bibfnamefont {H.}~\bibnamefont {Huang}}, \bibinfo
  {author} {\bibfnamefont {W.}~\bibnamefont {Jiang}}, \bibinfo {author}
  {\bibfnamefont {X.}~\bibnamefont {Ni}}, \bibinfo {author} {\bibfnamefont
  {L.}~\bibnamefont {Kang}}, \bibinfo {author} {\bibfnamefont {S.}~\bibnamefont
  {Zhang}}, \bibinfo {author} {\bibfnamefont {H.}~\bibnamefont {Xu}}, \bibinfo
  {author} {\bibfnamefont {Z.}~\bibnamefont {Liu}},\ and\ \bibinfo {author}
  {\bibfnamefont {F.}~\bibnamefont {Liu}},\ }\bibfield  {title} {\bibinfo
  {title} {Kagome bands disguised in a coloring-triangle lattice},\ }\href
  {https://doi.org/10.1103/PhysRevB.99.100404} {\bibfield  {journal} {\bibinfo
  {journal} {Phys. Rev. B}\ }\textbf {\bibinfo {volume} {99}},\ \bibinfo
  {pages} {100404} (\bibinfo {year} {2019})}\BibitemShut {NoStop}%
\bibitem [{\citenamefont {Graf}\ and\ \citenamefont
  {Pi{\'e}chon}(2021)}]{graf2021designing}%
  \BibitemOpen
  \bibfield  {author} {\bibinfo {author} {\bibfnamefont {A.}~\bibnamefont
  {Graf}}\ and\ \bibinfo {author} {\bibfnamefont {F.}~\bibnamefont
  {Pi{\'e}chon}},\ }\bibfield  {title} {\bibinfo {title} {Designing flat-band
  tight-binding models with tunable multifold band touching points},\ }\href
  {https://doi.org/https://doi.org/10.1103/PhysRevB.104.195128} {\bibfield
  {journal} {\bibinfo  {journal} {Phys. Rev. B}\ }\textbf {\bibinfo {volume}
  {104}},\ \bibinfo {pages} {195128} (\bibinfo {year} {2021})}\BibitemShut
  {NoStop}%
\bibitem [{\citenamefont {Jiang}\ \emph
  {et~al.}(2019{\natexlab{b}})\citenamefont {Jiang}, \citenamefont {Kang},
  \citenamefont {Huang}, \citenamefont {Xu}, \citenamefont {Low},\ and\
  \citenamefont {Liu}}]{jiang2019topological}%
  \BibitemOpen
  \bibfield  {author} {\bibinfo {author} {\bibfnamefont {W.}~\bibnamefont
  {Jiang}}, \bibinfo {author} {\bibfnamefont {M.}~\bibnamefont {Kang}},
  \bibinfo {author} {\bibfnamefont {H.}~\bibnamefont {Huang}}, \bibinfo
  {author} {\bibfnamefont {H.}~\bibnamefont {Xu}}, \bibinfo {author}
  {\bibfnamefont {T.}~\bibnamefont {Low}},\ and\ \bibinfo {author}
  {\bibfnamefont {F.}~\bibnamefont {Liu}},\ }\bibfield  {title} {\bibinfo
  {title} {Topological band evolution between lieb and kagome lattices},\
  }\href {https://doi.org/10.1103/PhysRevB.99.125131} {\bibfield  {journal}
  {\bibinfo  {journal} {Phys. Rev. B}\ }\textbf {\bibinfo {volume} {99}},\
  \bibinfo {pages} {125131} (\bibinfo {year} {2019}{\natexlab{b}})}\BibitemShut
  {NoStop}%
\bibitem [{\citenamefont {Haldane}(1988)}]{haldane1988model}%
  \BibitemOpen
  \bibfield  {author} {\bibinfo {author} {\bibfnamefont {F.~D.~M.}\
  \bibnamefont {Haldane}},\ }\bibfield  {title} {\bibinfo {title} {Model for a
  quantum hall effect without landau levels: Condensed-matter realization of
  the" parity anomaly"},\ }\href {https://doi.org/10.1103/PhysRevLett.61.2015}
  {\bibfield  {journal} {\bibinfo  {journal} {Phys. Rev. Lett.}\ }\textbf
  {\bibinfo {volume} {61}},\ \bibinfo {pages} {2015} (\bibinfo {year}
  {1988})}\BibitemShut {NoStop}%
\bibitem [{\citenamefont {Singh}\ \emph {et~al.}(2022)\citenamefont {Singh},
  \citenamefont {Lin},\ and\ \citenamefont {Bansil}}]{Singh2022}%
  \BibitemOpen
  \bibfield  {author} {\bibinfo {author} {\bibfnamefont {B.}~\bibnamefont
  {Singh}}, \bibinfo {author} {\bibfnamefont {H.}~\bibnamefont {Lin}},\ and\
  \bibinfo {author} {\bibfnamefont {A.}~\bibnamefont {Bansil}},\ }\bibfield
  {title} {\bibinfo {title} {Topology and symmetry in quantum materials},\
  }\href {https://doi.org/10.1002/adma.202201058} {\bibfield  {journal}
  {\bibinfo  {journal} {Adv. Mater.}\ }\textbf {\bibinfo {volume} {35}},\
  \bibinfo {pages} {2201058} (\bibinfo {year} {2022})}\BibitemShut {NoStop}%
\bibitem [{\citenamefont {Hasan}\ and\ \citenamefont {Kane}(2010)}]{hasan2010}%
  \BibitemOpen
  \bibfield  {author} {\bibinfo {author} {\bibfnamefont {M.~Z.}\ \bibnamefont
  {Hasan}}\ and\ \bibinfo {author} {\bibfnamefont {C.~L.}\ \bibnamefont
  {Kane}},\ }\bibfield  {title} {\bibinfo {title} {Colloquium: Topological
  insulators},\ }\href {https://doi.org/10.1103/RevModPhys.82.3045} {\bibfield
  {journal} {\bibinfo  {journal} {Rev. Mod. Phys.}\ }\textbf {\bibinfo {volume}
  {82}},\ \bibinfo {pages} {3045} (\bibinfo {year} {2010})}\BibitemShut
  {NoStop}%
\bibitem [{\citenamefont {Kane}\ and\ \citenamefont
  {Mele}(2005{\natexlab{b}})}]{kane2005}%
  \BibitemOpen
  \bibfield  {author} {\bibinfo {author} {\bibfnamefont {C.~L.}\ \bibnamefont
  {Kane}}\ and\ \bibinfo {author} {\bibfnamefont {E.~J.}\ \bibnamefont
  {Mele}},\ }\bibfield  {title} {\bibinfo {title} {Quantum spin hall effect in
  graphene},\ }\href {https://doi.org/10.1103/PhysRevLett.95.226801} {\bibfield
   {journal} {\bibinfo  {journal} {Phys. Rev. Lett.}\ }\textbf {\bibinfo
  {volume} {95}},\ \bibinfo {pages} {226801} (\bibinfo {year}
  {2005}{\natexlab{b}})}\BibitemShut {NoStop}%
\bibitem [{\citenamefont {He}\ \emph {et~al.}(2019)\citenamefont {He},
  \citenamefont {Ding}, \citenamefont {Zhou}, \citenamefont {Wang},\ and\
  \citenamefont {Gong}}]{he2019quasicrystalline}%
  \BibitemOpen
  \bibfield  {author} {\bibinfo {author} {\bibfnamefont {A.-L.}\ \bibnamefont
  {He}}, \bibinfo {author} {\bibfnamefont {L.-R.}\ \bibnamefont {Ding}},
  \bibinfo {author} {\bibfnamefont {Y.}~\bibnamefont {Zhou}}, \bibinfo {author}
  {\bibfnamefont {Y.-F.}\ \bibnamefont {Wang}},\ and\ \bibinfo {author}
  {\bibfnamefont {C.-D.}\ \bibnamefont {Gong}},\ }\bibfield  {title} {\bibinfo
  {title} {Quasicrystalline chern insulators},\ }\href
  {https://doi.org/10.1103/PhysRevB.100.214109} {\bibfield  {journal} {\bibinfo
   {journal} {Phys. Rev. B}\ }\textbf {\bibinfo {volume} {100}},\ \bibinfo
  {pages} {214109} (\bibinfo {year} {2019})}\BibitemShut {NoStop}%
\bibitem [{\citenamefont {Kraus}\ \emph {et~al.}(2012)\citenamefont {Kraus},
  \citenamefont {Lahini}, \citenamefont {Ringel}, \citenamefont {Verbin},\ and\
  \citenamefont {Zilberberg}}]{kraus2012}%
  \BibitemOpen
  \bibfield  {author} {\bibinfo {author} {\bibfnamefont {Y.~E.}\ \bibnamefont
  {Kraus}}, \bibinfo {author} {\bibfnamefont {Y.}~\bibnamefont {Lahini}},
  \bibinfo {author} {\bibfnamefont {Z.}~\bibnamefont {Ringel}}, \bibinfo
  {author} {\bibfnamefont {M.}~\bibnamefont {Verbin}},\ and\ \bibinfo {author}
  {\bibfnamefont {O.}~\bibnamefont {Zilberberg}},\ }\bibfield  {title}
  {\bibinfo {title} {Topological states and adiabatic pumping in
  quasicrystals},\ }\href {https://doi.org/10.1103/PhysRevLett.109.106402}
  {\bibfield  {journal} {\bibinfo  {journal} {Phys. Rev. Lett.}\ }\textbf
  {\bibinfo {volume} {109}},\ \bibinfo {pages} {106402} (\bibinfo {year}
  {2012})}\BibitemShut {NoStop}%
\bibitem [{\citenamefont {Hu}\ \emph {et~al.}(2011)\citenamefont {Hu},
  \citenamefont {Kargarian},\ and\ \citenamefont {Fiete}}]{hu2011topological}%
  \BibitemOpen
  \bibfield  {author} {\bibinfo {author} {\bibfnamefont {X.}~\bibnamefont
  {Hu}}, \bibinfo {author} {\bibfnamefont {M.}~\bibnamefont {Kargarian}},\ and\
  \bibinfo {author} {\bibfnamefont {G.~A.}\ \bibnamefont {Fiete}},\ }\bibfield
  {title} {\bibinfo {title} {Topological insulators and fractional quantum hall
  effect on the ruby lattice},\ }\href
  {https://doi.org/10.1103/PhysRevB.84.155116} {\bibfield  {journal} {\bibinfo
  {journal} {Phys. Rev. B}\ }\textbf {\bibinfo {volume} {84}},\ \bibinfo
  {pages} {155116} (\bibinfo {year} {2011})}\BibitemShut {NoStop}%
\bibitem [{\citenamefont {Sun}\ \emph {et~al.}(2011)\citenamefont {Sun},
  \citenamefont {Gu}, \citenamefont {Katsura},\ and\ \citenamefont
  {Das~Sarma}}]{sun2011nearly}%
  \BibitemOpen
  \bibfield  {author} {\bibinfo {author} {\bibfnamefont {K.}~\bibnamefont
  {Sun}}, \bibinfo {author} {\bibfnamefont {Z.}~\bibnamefont {Gu}}, \bibinfo
  {author} {\bibfnamefont {H.}~\bibnamefont {Katsura}},\ and\ \bibinfo {author}
  {\bibfnamefont {S.}~\bibnamefont {Das~Sarma}},\ }\bibfield  {title} {\bibinfo
  {title} {Nearly flatbands with nontrivial topology},\ }\href
  {https://doi.org/10.1103/PhysRevLett.106.236803} {\bibfield  {journal}
  {\bibinfo  {journal} {Phys. Rev. Lett.}\ }\textbf {\bibinfo {volume} {106}},\
  \bibinfo {pages} {236803} (\bibinfo {year} {2011})}\BibitemShut {NoStop}%
\bibitem [{\citenamefont {Wang}\ and\ \citenamefont
  {Ran}(2011)}]{wang2011nearly}%
  \BibitemOpen
  \bibfield  {author} {\bibinfo {author} {\bibfnamefont {F.}~\bibnamefont
  {Wang}}\ and\ \bibinfo {author} {\bibfnamefont {Y.}~\bibnamefont {Ran}},\
  }\bibfield  {title} {\bibinfo {title} {Nearly flat band with chern number c=
  2 on the dice lattice},\ }\href {https://doi.org/10.1103/PhysRevB.84.241103}
  {\bibfield  {journal} {\bibinfo  {journal} {Phys. Rev. B}\ }\textbf {\bibinfo
  {volume} {84}},\ \bibinfo {pages} {241103} (\bibinfo {year}
  {2011})}\BibitemShut {NoStop}%
\bibitem [{\citenamefont {Guan}\ \emph {et~al.}(2023)\citenamefont {Guan},
  \citenamefont {Qi}, \citenamefont {Zhang}, \citenamefont {Liu},\ and\
  \citenamefont {He}}]{guan2023staggered}%
  \BibitemOpen
  \bibfield  {author} {\bibinfo {author} {\bibfnamefont {D.-H.}\ \bibnamefont
  {Guan}}, \bibinfo {author} {\bibfnamefont {L.}~\bibnamefont {Qi}}, \bibinfo
  {author} {\bibfnamefont {X.}~\bibnamefont {Zhang}}, \bibinfo {author}
  {\bibfnamefont {Y.}~\bibnamefont {Liu}},\ and\ \bibinfo {author}
  {\bibfnamefont {A.-L.}\ \bibnamefont {He}},\ }\bibfield  {title} {\bibinfo
  {title} {Staggered magnetic flux induced higher-order topological insulators
  and topological flat bands on the ruby lattice},\ }\href
  {https://doi.org/https://doi.org/10.1103/PhysRevB.108.085121} {\bibfield
  {journal} {\bibinfo  {journal} {Phys. Rev. B}\ }\textbf {\bibinfo {volume}
  {108}},\ \bibinfo {pages} {085121} (\bibinfo {year} {2023})}\BibitemShut
  {NoStop}%
\bibitem [{\citenamefont {He}\ \emph {et~al.}(2024)\citenamefont {He},
  \citenamefont {Yan}, \citenamefont {Qi}, \citenamefont {Liu},\ and\
  \citenamefont {Han}}]{he2024topological}%
  \BibitemOpen
  \bibfield  {author} {\bibinfo {author} {\bibfnamefont {A.-L.}\ \bibnamefont
  {He}}, \bibinfo {author} {\bibfnamefont {X.-H.}\ \bibnamefont {Yan}},
  \bibinfo {author} {\bibfnamefont {L.}~\bibnamefont {Qi}}, \bibinfo {author}
  {\bibfnamefont {Y.}~\bibnamefont {Liu}},\ and\ \bibinfo {author}
  {\bibfnamefont {Y.}~\bibnamefont {Han}},\ }\bibfield  {title} {\bibinfo
  {title} {Topological states and flat bands on the maple leaf lattice},\
  }\href {https://doi.org/10.1103/PhysRevB.109.075118} {\bibfield  {journal}
  {\bibinfo  {journal} {Phys. Rev. B}\ }\textbf {\bibinfo {volume} {109}},\
  \bibinfo {pages} {075118} (\bibinfo {year} {2024})}\BibitemShut {NoStop}%
\bibitem [{\citenamefont {He}\ \emph {et~al.}(2022{\natexlab{a}})\citenamefont
  {He}, \citenamefont {Luo}, \citenamefont {Zhou}, \citenamefont {Wang},\ and\
  \citenamefont {Yao}}]{he2022topological}%
  \BibitemOpen
  \bibfield  {author} {\bibinfo {author} {\bibfnamefont {A.-L.}\ \bibnamefont
  {He}}, \bibinfo {author} {\bibfnamefont {W.-W.}\ \bibnamefont {Luo}},
  \bibinfo {author} {\bibfnamefont {Y.}~\bibnamefont {Zhou}}, \bibinfo {author}
  {\bibfnamefont {Y.-F.}\ \bibnamefont {Wang}},\ and\ \bibinfo {author}
  {\bibfnamefont {H.}~\bibnamefont {Yao}},\ }\bibfield  {title} {\bibinfo
  {title} {Topological states in a dimerized system with staggered magnetic
  fluxes},\ }\href {https://doi.org/10.1103/PhysRevB.106.125147} {\bibfield
  {journal} {\bibinfo  {journal} {Phys. Rev. B}\ }\textbf {\bibinfo {volume}
  {105}},\ \bibinfo {pages} {235139} (\bibinfo {year}
  {2022}{\natexlab{a}})}\BibitemShut {NoStop}%
\bibitem [{\citenamefont {Benalcazar}\ \emph {et~al.}(2017)\citenamefont
  {Benalcazar}, \citenamefont {Bernevig},\ and\ \citenamefont
  {Hughes}}]{benalcazar2017electric}%
  \BibitemOpen
  \bibfield  {author} {\bibinfo {author} {\bibfnamefont {W.~A.}\ \bibnamefont
  {Benalcazar}}, \bibinfo {author} {\bibfnamefont {B.~A.}\ \bibnamefont
  {Bernevig}},\ and\ \bibinfo {author} {\bibfnamefont {T.~L.}\ \bibnamefont
  {Hughes}},\ }\bibfield  {title} {\bibinfo {title} {Electric multipole
  moments, topological multipole moment pumping, and chiral hinge states in
  crystalline insulators},\ }\href {https://doi.org/10.1103/PhysRevB.96.245115}
  {\bibfield  {journal} {\bibinfo  {journal} {Phys. Rev. B}\ }\textbf {\bibinfo
  {volume} {96}},\ \bibinfo {pages} {245115} (\bibinfo {year}
  {2017})}\BibitemShut {NoStop}%
\bibitem [{\citenamefont {Liu}\ and\ \citenamefont
  {Wakabayashi}(2017)}]{liu2017novel}%
  \BibitemOpen
  \bibfield  {author} {\bibinfo {author} {\bibfnamefont {F.}~\bibnamefont
  {Liu}}\ and\ \bibinfo {author} {\bibfnamefont {K.}~\bibnamefont
  {Wakabayashi}},\ }\bibfield  {title} {\bibinfo {title} {Novel topological
  phase with a zero berry curvature},\ }\href
  {https://doi.org/10.1103/PhysRevLett.118.076803} {\bibfield  {journal}
  {\bibinfo  {journal} {Phys. Rev. Lett.}\ }\textbf {\bibinfo {volume} {118}},\
  \bibinfo {pages} {076803} (\bibinfo {year} {2017})}\BibitemShut {NoStop}%
\bibitem [{\citenamefont {Schindler}\ \emph {et~al.}(2018)\citenamefont
  {Schindler}, \citenamefont {Cook}, \citenamefont {Vergniory}, \citenamefont
  {Wang}, \citenamefont {Parkin}, \citenamefont {Bernevig},\ and\ \citenamefont
  {Neupert}}]{schindler2018higher}%
  \BibitemOpen
  \bibfield  {author} {\bibinfo {author} {\bibfnamefont {F.}~\bibnamefont
  {Schindler}}, \bibinfo {author} {\bibfnamefont {A.~M.}\ \bibnamefont {Cook}},
  \bibinfo {author} {\bibfnamefont {M.~G.}\ \bibnamefont {Vergniory}}, \bibinfo
  {author} {\bibfnamefont {Z.}~\bibnamefont {Wang}}, \bibinfo {author}
  {\bibfnamefont {S.~S.}\ \bibnamefont {Parkin}}, \bibinfo {author}
  {\bibfnamefont {B.~A.}\ \bibnamefont {Bernevig}},\ and\ \bibinfo {author}
  {\bibfnamefont {T.}~\bibnamefont {Neupert}},\ }\bibfield  {title} {\bibinfo
  {title} {Higher-order topological insulators},\ }\href
  {https://doi.org/10.1126/sciadv.aat0346} {\bibfield  {journal} {\bibinfo
  {journal} {Sci. Adv.}\ }\textbf {\bibinfo {volume} {4}},\ \bibinfo {pages}
  {eaat0346} (\bibinfo {year} {2018})}\BibitemShut {NoStop}%
\bibitem [{\citenamefont {Franca}\ \emph {et~al.}(2018)\citenamefont {Franca},
  \citenamefont {van~den Brink},\ and\ \citenamefont
  {Fulga}}]{franca2018anomalous}%
  \BibitemOpen
  \bibfield  {author} {\bibinfo {author} {\bibfnamefont {S.}~\bibnamefont
  {Franca}}, \bibinfo {author} {\bibfnamefont {J.}~\bibnamefont {van~den
  Brink}},\ and\ \bibinfo {author} {\bibfnamefont {I.}~\bibnamefont {Fulga}},\
  }\bibfield  {title} {\bibinfo {title} {An anomalous higher-order topological
  insulator},\ }\href {https://doi.org/10.1103/PhysRevB.98.201114} {\bibfield
  {journal} {\bibinfo  {journal} {Phys. Rev. B}\ }\textbf {\bibinfo {volume}
  {98}},\ \bibinfo {pages} {201114} (\bibinfo {year} {2018})}\BibitemShut
  {NoStop}%
\bibitem [{\citenamefont {Liu}\ \emph {et~al.}(2019)\citenamefont {Liu},
  \citenamefont {Deng},\ and\ \citenamefont {Wakabayashi}}]{liu2019helical}%
  \BibitemOpen
  \bibfield  {author} {\bibinfo {author} {\bibfnamefont {F.}~\bibnamefont
  {Liu}}, \bibinfo {author} {\bibfnamefont {H.-Y.}\ \bibnamefont {Deng}},\ and\
  \bibinfo {author} {\bibfnamefont {K.}~\bibnamefont {Wakabayashi}},\
  }\bibfield  {title} {\bibinfo {title} {Helical topological edge states in a
  quadrupole phase},\ }\href {https://doi.org/10.1103/PhysRevLett.122.086804}
  {\bibfield  {journal} {\bibinfo  {journal} {Phys. Rev. Lett.}\ }\textbf
  {\bibinfo {volume} {122}},\ \bibinfo {pages} {086804} (\bibinfo {year}
  {2019})}\BibitemShut {NoStop}%
\bibitem [{\citenamefont {Kargarian}\ and\ \citenamefont
  {Fiete}(2010)}]{kargarian2010topological}%
  \BibitemOpen
  \bibfield  {author} {\bibinfo {author} {\bibfnamefont {M.}~\bibnamefont
  {Kargarian}}\ and\ \bibinfo {author} {\bibfnamefont {G.~A.}\ \bibnamefont
  {Fiete}},\ }\bibfield  {title} {\bibinfo {title} {Topological phases and
  phase transitions on the square-octagon lattice},\ }\href
  {https://doi.org/10.1103/PhysRevB.82.085106} {\bibfield  {journal} {\bibinfo
  {journal} {Phys. Rev. B}\ }\textbf {\bibinfo {volume} {82}},\ \bibinfo
  {pages} {085106} (\bibinfo {year} {2010})}\BibitemShut {NoStop}%
\bibitem [{\citenamefont {Pal}(2018)}]{pal2018nontrivial}%
  \BibitemOpen
  \bibfield  {author} {\bibinfo {author} {\bibfnamefont {B.}~\bibnamefont
  {Pal}},\ }\bibfield  {title} {\bibinfo {title} {Nontrivial topological flat
  bands in a diamond-octagon lattice geometry},\ }\href
  {https://doi.org/10.1103/PhysRevB.98.245116} {\bibfield  {journal} {\bibinfo
  {journal} {Phys. Rev. B}\ }\textbf {\bibinfo {volume} {98}},\ \bibinfo
  {pages} {245116} (\bibinfo {year} {2018})}\BibitemShut {NoStop}%
\bibitem [{\citenamefont {Nunes}\ and\ \citenamefont
  {Smith}(2020)}]{nunes2020flat}%
  \BibitemOpen
  \bibfield  {author} {\bibinfo {author} {\bibfnamefont {L.~H.}\ \bibnamefont
  {Nunes}}\ and\ \bibinfo {author} {\bibfnamefont {C.~M.}\ \bibnamefont
  {Smith}},\ }\bibfield  {title} {\bibinfo {title} {Flat-band superconductivity
  for tight-binding electrons on a square-octagon lattice},\ }\href
  {https://doi.org/10.1103/PhysRevB.101.224514} {\bibfield  {journal} {\bibinfo
   {journal} {Phys. Rev. B}\ }\textbf {\bibinfo {volume} {101}},\ \bibinfo
  {pages} {224514} (\bibinfo {year} {2020})}\BibitemShut {NoStop}%
\bibitem [{\citenamefont {He}\ \emph {et~al.}(2022{\natexlab{b}})\citenamefont
  {He}, \citenamefont {Zhang},\ and\ \citenamefont {Liu}}]{he2022topological1}%
  \BibitemOpen
  \bibfield  {author} {\bibinfo {author} {\bibfnamefont {A.-L.}\ \bibnamefont
  {He}}, \bibinfo {author} {\bibfnamefont {X.}~\bibnamefont {Zhang}},\ and\
  \bibinfo {author} {\bibfnamefont {Y.}~\bibnamefont {Liu}},\ }\bibfield
  {title} {\bibinfo {title} {Topological states in a dimerized square-octagon
  lattice with staggered magnetic fluxes},\ }\href
  {https://doi.org/10.1103/PhysRevB.106.125147} {\bibfield  {journal} {\bibinfo
   {journal} {Phys. Rev. B}\ }\textbf {\bibinfo {volume} {106}},\ \bibinfo
  {pages} {125147} (\bibinfo {year} {2022}{\natexlab{b}})}\BibitemShut
  {NoStop}%
\bibitem [{\citenamefont {Yan}\ \emph {et~al.}(2023)\citenamefont {Yan},
  \citenamefont {Zhang}, \citenamefont {Wang},\ and\ \citenamefont
  {Yan}}]{yan2023intrinsic}%
  \BibitemOpen
  \bibfield  {author} {\bibinfo {author} {\bibfnamefont {L.}~\bibnamefont
  {Yan}}, \bibinfo {author} {\bibfnamefont {D.}~\bibnamefont {Zhang}}, \bibinfo
  {author} {\bibfnamefont {X.-J.}\ \bibnamefont {Wang}},\ and\ \bibinfo
  {author} {\bibfnamefont {J.-Y.}\ \bibnamefont {Yan}},\ }\bibfield  {title}
  {\bibinfo {title} {Intrinsic topological metal state in t-graphene},\ }\href
  {https://doi.org/10.1088/1367-2630/acccd7} {\bibfield  {journal} {\bibinfo
  {journal} {New J. Phys.}\ }\textbf {\bibinfo {volume} {25}},\ \bibinfo
  {pages} {043020} (\bibinfo {year} {2023})}\BibitemShut {NoStop}%
\bibitem [{\citenamefont {He}\ and\ \citenamefont {Liu}(2023)}]{he2023dirac}%
  \BibitemOpen
  \bibfield  {author} {\bibinfo {author} {\bibfnamefont {J.}~\bibnamefont
  {He}}\ and\ \bibinfo {author} {\bibfnamefont {Z.}~\bibnamefont {Liu}},\
  }\bibfield  {title} {\bibinfo {title} {Dirac cones in bipartite
  square--octagon lattice: A theoretical approach},\ }\href
  {https://doi.org/10.1063/5.0160658} {\bibfield  {journal} {\bibinfo
  {journal} {J. Chem. Phys.}\ }\textbf {\bibinfo {volume} {159}},\ \bibinfo
  {pages} {044713} (\bibinfo {year} {2023})}\BibitemShut {NoStop}%
\bibitem [{\citenamefont {Thouless}\ \emph {et~al.}(1982)\citenamefont
  {Thouless}, \citenamefont {Kohmoto}, \citenamefont {Nightingale},\ and\
  \citenamefont {den Nijs}}]{thouless1982quantized}%
  \BibitemOpen
  \bibfield  {author} {\bibinfo {author} {\bibfnamefont {D.~J.}\ \bibnamefont
  {Thouless}}, \bibinfo {author} {\bibfnamefont {M.}~\bibnamefont {Kohmoto}},
  \bibinfo {author} {\bibfnamefont {M.~P.}\ \bibnamefont {Nightingale}},\ and\
  \bibinfo {author} {\bibfnamefont {M.}~\bibnamefont {den Nijs}},\ }\bibfield
  {title} {\bibinfo {title} {Quantized hall conductance in a two-dimensional
  periodic potential},\ }\href {https://doi.org/10.1103/PhysRevLett.49.405}
  {\bibfield  {journal} {\bibinfo  {journal} {Phys. Rev. Lett.}\ }\textbf
  {\bibinfo {volume} {49}},\ \bibinfo {pages} {405} (\bibinfo {year}
  {1982})}\BibitemShut {NoStop}%
\bibitem [{\citenamefont {Fu}\ and\ \citenamefont
  {Kane}(2007)}]{fu2007topological}%
  \BibitemOpen
  \bibfield  {author} {\bibinfo {author} {\bibfnamefont {L.}~\bibnamefont
  {Fu}}\ and\ \bibinfo {author} {\bibfnamefont {C.~L.}\ \bibnamefont {Kane}},\
  }\bibfield  {title} {\bibinfo {title} {Topological insulators with inversion
  symmetry},\ }\href {https://doi.org/10.1103/PhysRevB.76.045302} {\bibfield
  {journal} {\bibinfo  {journal} {Phys. Rev. B}\ }\textbf {\bibinfo {volume}
  {76}},\ \bibinfo {pages} {045302} (\bibinfo {year} {2007})}\BibitemShut
  {NoStop}%
\bibitem [{\citenamefont {Hughes}\ \emph {et~al.}(2011)\citenamefont {Hughes},
  \citenamefont {Prodan},\ and\ \citenamefont
  {Bernevig}}]{hughes2011inversion}%
  \BibitemOpen
  \bibfield  {author} {\bibinfo {author} {\bibfnamefont {T.~L.}\ \bibnamefont
  {Hughes}}, \bibinfo {author} {\bibfnamefont {E.}~\bibnamefont {Prodan}},\
  and\ \bibinfo {author} {\bibfnamefont {B.~A.}\ \bibnamefont {Bernevig}},\
  }\bibfield  {title} {\bibinfo {title} {Inversion-symmetric topological
  insulators},\ }\href {https://doi.org/10.1103/PhysRevB.83.245132} {\bibfield
  {journal} {\bibinfo  {journal} {Phys. Rev. B}\ }\textbf {\bibinfo {volume}
  {83}},\ \bibinfo {pages} {245132} (\bibinfo {year} {2011})}\BibitemShut
  {NoStop}%
\bibitem [{\citenamefont {Fang}\ \emph {et~al.}(2012)\citenamefont {Fang},
  \citenamefont {Gilbert},\ and\ \citenamefont {Bernevig}}]{fang2012bulk}%
  \BibitemOpen
  \bibfield  {author} {\bibinfo {author} {\bibfnamefont {C.}~\bibnamefont
  {Fang}}, \bibinfo {author} {\bibfnamefont {M.~J.}\ \bibnamefont {Gilbert}},\
  and\ \bibinfo {author} {\bibfnamefont {B.~A.}\ \bibnamefont {Bernevig}},\
  }\bibfield  {title} {\bibinfo {title} {Bulk topological invariants in
  noninteracting point group symmetric insulators},\ }\href
  {https://doi.org/10.1103/PhysRevB.86.115112} {\bibfield  {journal} {\bibinfo
  {journal} {Phys. Rev. B}\ }\textbf {\bibinfo {volume} {86}},\ \bibinfo
  {pages} {115112} (\bibinfo {year} {2012})}\BibitemShut {NoStop}%
\end{thebibliography}%
\end{document}